\documentclass[twocolumn, bibnotes, nofootinbib, showpacs,preprintnumbers,amsmath,amssymb]{revtex4-1}
\usepackage{graphicx}
\usepackage{epsf}
\usepackage{amsmath}
\usepackage{amsfonts}
\usepackage{amssymb}
\usepackage{hyperref}

\begin{document}
\preprint{APS}

\title{Onset of Singularities in the Pattern of Fluctuational Paths of a Nonequilibrium System.}
\author{Oleg Kogan}
\email{obk5@cornell.edu}
\affiliation{Laboratory of Atomic and Solid State Physics, Cornell University, Ithaca, NY, 14853}

\date{\today}

\begin{abstract}
Fluctuations in systems away from thermal equilibrium have features that have no analog in equilibrium systems. One of such features concerns large rare excursions far from the stable state in the space of dynamical variables. For equilibrium systems, the most probable fluctuational trajectory to a given state is related to the fluctuation-free trajectory back to the stable state by time reversal. This is no longer true for nonequilibrium systems, where the pattern of the most probable trajectories generally displays singularities. Here we study how the singularities emerge as the system is driven away from equilibrium, and whether a driving strength threshold is required for their onset.  Using a resonantly modulated oscillator as a model, we identify two distinct scenarios, depending on the speed of the optimal path in thermal equilibrium.  If the position away from the stable state along the optimal path grows exponentially in time, the singularities emerge without a threshold.  We find the scaling of the location of the singularities as a function of the control parameter.  If the growth away from the stable state is faster than exponential, characterized by the ability to reach infinity in finite time,  there is a threshold for the onset of singularities, which we study for the model.

\end{abstract}

\maketitle

\section{Introduction}
The phenomena of large rare fluctuations have been attracting increasing attention recently in the context of small systems, such as nanomechanical resonators \cite{Nanomechanics1, Nanomechanics2}, nanomagnets \cite{Nanomagnets, Nanomagnets2}, Josephson junctions \cite{Josephson}, and trapped ions \cite{Tony}.  Another arena of rare events is biology.  Despite being inherently stochastic \cite{Bio 1}, biological processes are notoriously robust \cite{Bio 2, Bio 3}, yet prone to mutations, and hence to disease, genetic variability, evolution, as well as making functional use of the noise \cite{Bio 4}.   Large fluctuations may also manifest themselves in catastrophic events in various phenomena, such as in climate \cite{Climate}, or population dynamics \cite{Population_bio}.

The problem of fluctuations has two closely related aspects: the form of the probability distribution and properties of paths in the dynamical configuration space followed during a fluctuation.  Both aspects are well-understood for systems in thermal equilibrium.  Here the probability density distribution is given by the Boltzmann law.  When the fluctuations caused by thermal noise are weak on average, this distribution is narrowly peaked at the stable state of the system in phase space and usually rapidly falls down away from it.  The paths followed in fluctuations were first studied by Onsager and Machlup \cite{OM1,OM2}, who showed that in a large fluctuation to a remote part of the phase space the system is most likely to follow a certain optimal path.  For systems in thermal equilibrium this path is simply related to the fluctuation-free path (may also be called relaxational or mean-field path) back to the stable state via a time-reversal relationship.  This relationship follows from the principle of maximal work for equilibrium systems \cite{LL}.  
It implies that if the pattern of relaxational paths is unique, i.e. relaxation from a given point to an attractor is accomplished via a unique path (this is a generic situation), then the pattern of optimal fluctuational paths, without conditioning on the fluctuation time, will also be unique - a given point in phase space is reached from an attracting state by a unique fluctuational path.  In contrast, in nonequilibrium systems there is no simple relation between the most probable fluctuational and relaxational paths.  Moreover, generically the pattern of optimal fluctuational paths may allow for intersection of neighboring paths, and the loci of points in phase space along which this happens are the caustic singulatiries \cite{Schulman, Mark and Marko}.  Such singularities are a counterpart of caustics in optics \cite{Optics_caustics} and in the WKB trajectories in Quantum Mechanics \cite{QM_WKB}.  Caustics are accompanied by switching lines which separate regions of phase space such that the neighboring points on the opposite sides of switching lines will be reached by non-neighboring fluctuational paths.  The gradient of the probability distribution has discontinuities along switching lines.  Switching lines are the experimentally-observable signatures of singularities.  The existence of singularities implies a stark qualitative difference from equilibrium - both in the pattern of optimal paths and in the form of the probability distribution.  

Such picture of the structure of singularities is understood \cite{Mark and Marko}.  The separate question of the mechanism by which they are formed has been studied for systems with a saddle point by Dykman, Millonas, and Smelyanskiy \cite{Mark and Marko} (see also \cite{Mark_chemistry}) and by Maier and Stein (for example in \cite{MS_focusing-caustics}, among others), and for systems with an unstable limit cycle in \cite{Limit Cycle}.  Here we focus specifically on the question of threshold in a monostable system as it is driven away from equilibrium.  We frame the study of singularities of optimal paths in the following context.  Consider a generic system with a finite-dimensional phase space.  Driving of a system from equilibrium will be characterized by a certain control parameter that serves as a quantitative measure of the deviation from equilibrium.  Singularities will generically appear at some critical value of this control parameter.  With this picture in mind, we formulate our main question as follows: what properties of the system determine whether this critical strength of the driving is finite or infinitesimal?  The effect of the perturbation to first-order is computed along the unperturbed path, and it is given by a certain functional of the unperturbed path.  Therefore, the answer to our main question depends on the form of the perturbation and on the properties of unperturbed paths.  Focusing on the case of non-singular perturbations, we point out the qualitative difference between those unperturbed paths that reach infinity in an infinite or finite time, respectively giving rise to a zero versus a finite critical driving threshold at which singularities appear.

In this paper we study how singularities in the pattern of optimal paths emerge in a particular system as it is driven away from thermal equilibrium - a nonlinear oscillator driven by periodic field and interacting with a thermal bath.  
The amplitude of the periodic driving field is a control parameter that serves as a quantitative measure of the deviation from equilibrium.  The situation at zero field is straightforward and agrees with predictions of equilibrium statistical mechanics: the probability distribution in the phase space of the oscillator is predictably Boltzman and a most-likely fluctuation to any specified point of phase space is realized via a unique path.  Moreover, these paths can be obtained from the noise-free paths via a time-reversal relationship. 
In contrast, the situation at finite driving is far less clear.  We focus on the question of the driving threshold above which the singularities develop - is it finite or infinitesimal, and what properties of the system control it?  

In our oscillator case-study, the presence of finite or infinitesimal driving threshold turns out to depend on the strength of the nonlinear damping.  When the
nonlinear damping is present, the value of the threshold driving field required for the appearance of caustics is nonzero.  This threshold
field scales linearly with the strength of the nonlinear damping and becomes zero when only the linear damping is present.  In this regime of purely linear damping, caustics appear at an infinitesimal value of the driving field.  This is a very interesting scenario: the system becomes qualitatively different from an equilibrium one even if it is driven away from
equilibrium by an infinitesimal amount - a type of an equilibrium fragility.
Secondly, we found that 
as the driving field passes above the threshold, the caustics arrive from infinity and move in closer to the attracting stable state as the driving field increases.  
We used perturbation theory to understand the mechanism of caustic formation.  We learned that caustics form as a result of the skewness (angular asymetry) of the pattern of fluctuational paths.  This skewness decreases with decreasing driving field, and this is the reason why caustics are found further away from the stable state as driving goes to zero.  
The analytical treatment also explains the appearance of a finite threshold when nonlinear damping is finite.  The presence of nonlinear damping causes faster than exponential dynamics and therefore allows paths to reach infinity in finite time.  The interplay between the skewness, which tends to cause caustics, and this accelerated motion in the presence of nonlinear damping is responsible for the threshold.  We emphasize that the ability to reach infinity in finite or infinite time serves merely as a way to separate the speed of optimal paths into two distinct categories.  The physics does not depend on the point at infinity, but it does depend on the speed of the optimal paths in the relevant parts of phase space.  In fact, the model needs to be modified far away from the stable state if it is to remain realistic.  However, as the driving field grows, the caustics are situated sufficiently close to the stable state where the present model is realistic. 

This paper is organized as follows.  Section \ref{sec:Phenomenology} is devoted to the exposition of the phenomenology of singular features.  In Section \ref{sec:Analytics} we outline the analytical approach and compare it with numerical predictions.  We further discuss the relevance of this work in Section \ref{sec:Discussion}.

\section{Pattern of Fluctuational Paths in the Phase Space of a Nonlinear Oscillator}
\label{sec:Phenomenology}
\subsection{Linear Damping}
\label{sec:Linear_Damping_Analysis}
We consider a classical nonlinear oscillator coupled to a heat bath
\begin{equation}
\label{eq:Langevin_linear}
\ddot{q} + \gamma \dot{q} + \omega_0^2 q + \alpha q^3 = F\cos{\left(\Omega t\right)} + f(t)
\end{equation}
where the effect of the bath has been reduced to the damping and noise terms which satisfy a fluctuation-dissipation relation with temperature $\tau$,
\begin{equation}
\langle f(t)f(t') \rangle = 2 \gamma k_B\tau  \delta(t-t').
\end{equation}
This description can be obtained, for example, by treating the bath as a collection of harmonic oscillators, each coupled linearly to the oscillator described by the variable $q$ \cite{Mark Review}.  In the regime when the resonance is sharp: $\gamma/\omega_0 \ll 1$ and the oscillator is driven close to resonance: $\left|\Omega - \omega_0 \right| \sim \gamma$, its dynamics is given approximately by fast simple harmonic motion modulated by a slow function of time, which is described by the following amplitude equations in ``slow time'' $T$ (see Appendix A) :
\begin{eqnarray}
\label{eq:AmpEq_linearQ}
\frac{d Q}{dT} &=& K_Q(Q,P)  + f_Q(T)  \\ 
\label{eq:AmpEq_linearP} 
\frac{d P}{dT} &=& K_P(Q,P) + f_P(T) 
\end{eqnarray}
where
\begin{eqnarray}
\label{eq:linearKQ}
K_Q &=&  \frac{\partial g}{\partial P} - \eta Q\\
\label{eq:linearKP}
K_P &=&  - \frac{\partial g}{\partial Q} - \eta P \\
\label{eq:g}
g &=& \frac{1}{4}\left( Q^2 +P^2 -1\right)^2 - Q\mathcal{F} 
\end{eqnarray}
\begin{equation}
\label{eq:SlowNoise_CorrFn}
\langle f_i(T)f_j(T') \rangle = D \delta_{i,j} \delta(T-T'),
\end{equation}
which can all be obtained from the microscopic Hamiltonian 
by integrating out the bath degrees of freedom \cite{Mark Review}  or from the corresponding Langevin equation, Eq.~(\ref{eq:Langevin_linear}) \cite{Thesis, Nayfeh, B&M}.   
The functional probability density of a given realization of a fluctuation can be expressed as
\begin{equation}
p(path) \sim \exp{\left(-\frac{S\left[f_Q(t),f_P(t),\lambda_Q(t) \lambda_P(t), Q(t),P(t)\right]}{D}\right)}. 
\end{equation}
For Gaussian noise with the correlator $\langle f_i(t)f_j(t') \rangle = D \delta_{i,j} \delta(t-t')$ (white noise), the functional $S$ is given to lowest order in $D$ by  \cite{Path_Integral_1, Path_Integral_2, Feynman and Hibbs}:
\begin{equation}
\label{eq:S_linear}
S  = \sum_{j=\{Q,P\}} \int_{-\infty}^{\infty} \left[\frac{1}{4}f_j^2(t) -  i \lambda_j(t)\left(\dot{x}_j - K_j - f_j(t) \right)\right] \,dt 
\end{equation}
where $x$ stands for either $Q(t)$ or $P(t)$ and $K$s are also functions of these dynamical variables.   Performing the functional integral in $\lambda$s and $f$s  
leaving the following Wentzel-Freidlin action \cite{Wentzel-Freidlin}.
\begin{equation}
\label{eq:WF_action}
S = \frac{1}{4} \int_{-\infty}^{\infty} \left[\left(\dot{Q} - K_Q\right)^2 + \left(\dot{P} - K_P\right)^2\right]  \,dt.
\end{equation}
Because our goal in this work is to understand the properties of the exponential part of distributions in the limit of weak noise strength $D$ we follow the WKB approach - evaluate the functional along the paths which minimize $S$, or optimal paths.  The problem then boils down do to finding optimal $Q(t)$ and $P(t)$ which minimize the associated action $S$.   Interpreting the integrand in Eq.~(\ref{eq:WF_action}) as the Lagrangian, the associated Wentzel-Freidlin auxiliary Hamiltonian is
\begin{equation}
\mathcal{H} = p_Q^2 + p_P^2 + p_QK_Q + p_P K_P, 
\end{equation}
and the optimal paths are solutions to the corresponding Hamiltonian motion generated by this auxiliary $\mathcal{H}$.  We will now study the properties of these trajectories numerically, and then analytically, with the help of the perturbation theory in Section \ref{sec:Analytics_linear}.
\subsubsection{Numerical procedure}
\label{sec:num_proc}
We note that $\vec{p} = \frac{1}{2}(\vec{\dot{x}} - \vec{K}) = $ noise, so in the absence
of noise, the only possible dynamics lie on the $\vec{p} = 0$
plane - the ``mean-field'' relaxational dynamics.  Any fluctuational dynamics
will be lifted out of this plane.  The fixed
points (FP) of this dynamics are $(Q_{FP},P_{FP},0,0)$ where
$(Q_{FP},P_{FP})$ are the FP of the original 2D dynamics.  It can also
be easily shown that eigenvalues of each of these auxiliary FP are
$(\lambda_1, \lambda_2, -\lambda_1, -\lambda_2)$ where $\lambda_1$
and $\lambda_2$ are eigenvalues of the FP of the original
dynamics. The $\vec{p} = 0$ plane forms the stable manifold of
each of the fixed points.  A trajectory that escapes from an attracting fixed point will lie on a certain manifold in the auxiliary $(Q,P,p_Q,
p_P)$ space.  In the vicinity of the attracting fixed point, this
manifold is tangent to the plane spanned by two unstable
eigenvectors of this FP.  Thus, the fluctuational trajectories lie on the unstable ``Lagrangian Manifold'' \cite{Graham DB, MS SIAM, Lagrangian_Manifolds, Arnold} of the attracting fixed point.  We calculate the set of trajectories that
escape the attractor by evolving a set of initial conditions placed on a circle on this unstable manifold very close to the attracting FP.   

\begin{figure}
\begin{center}
\includegraphics[width=12cm]{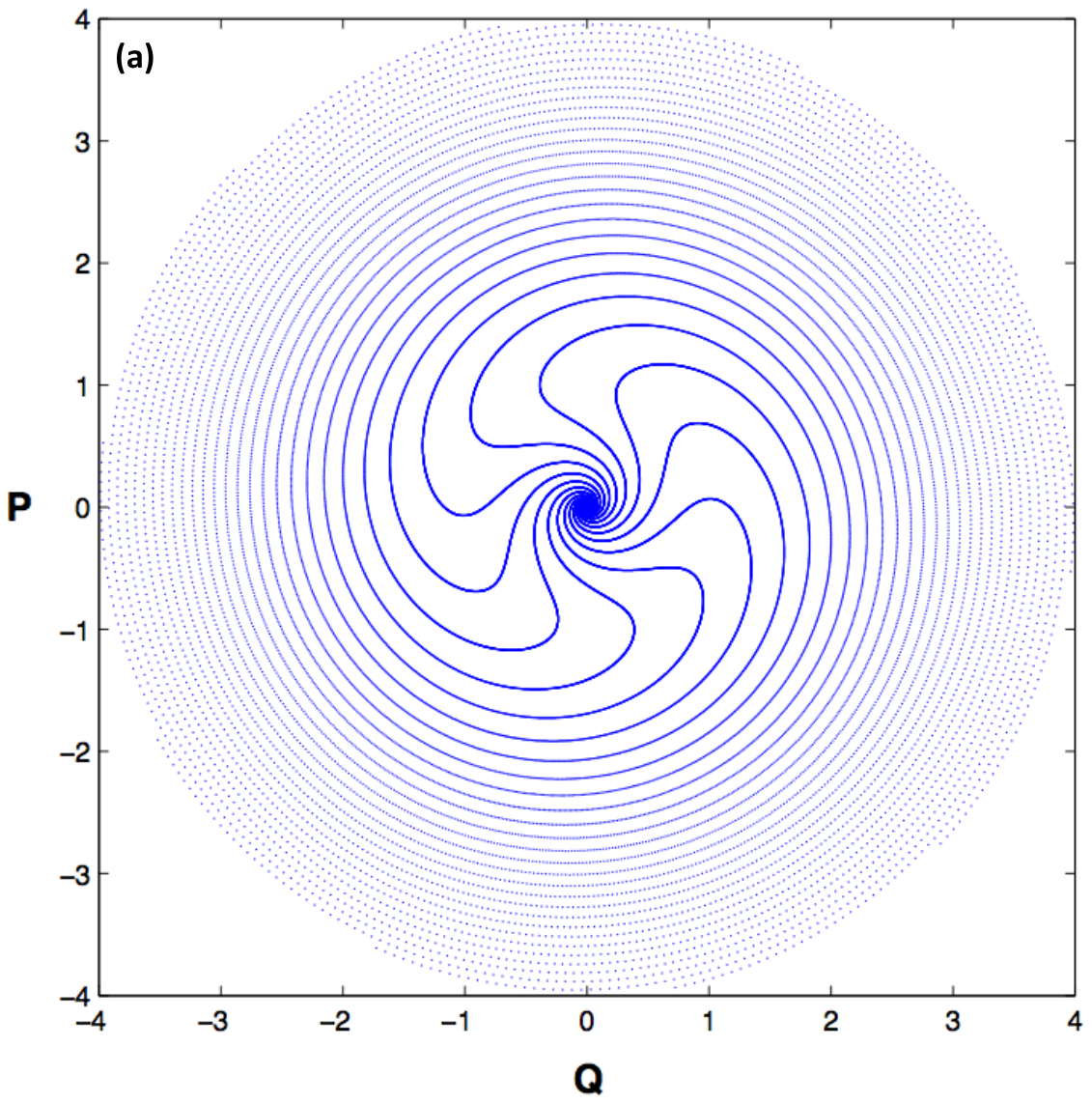}
\includegraphics[width=12cm]{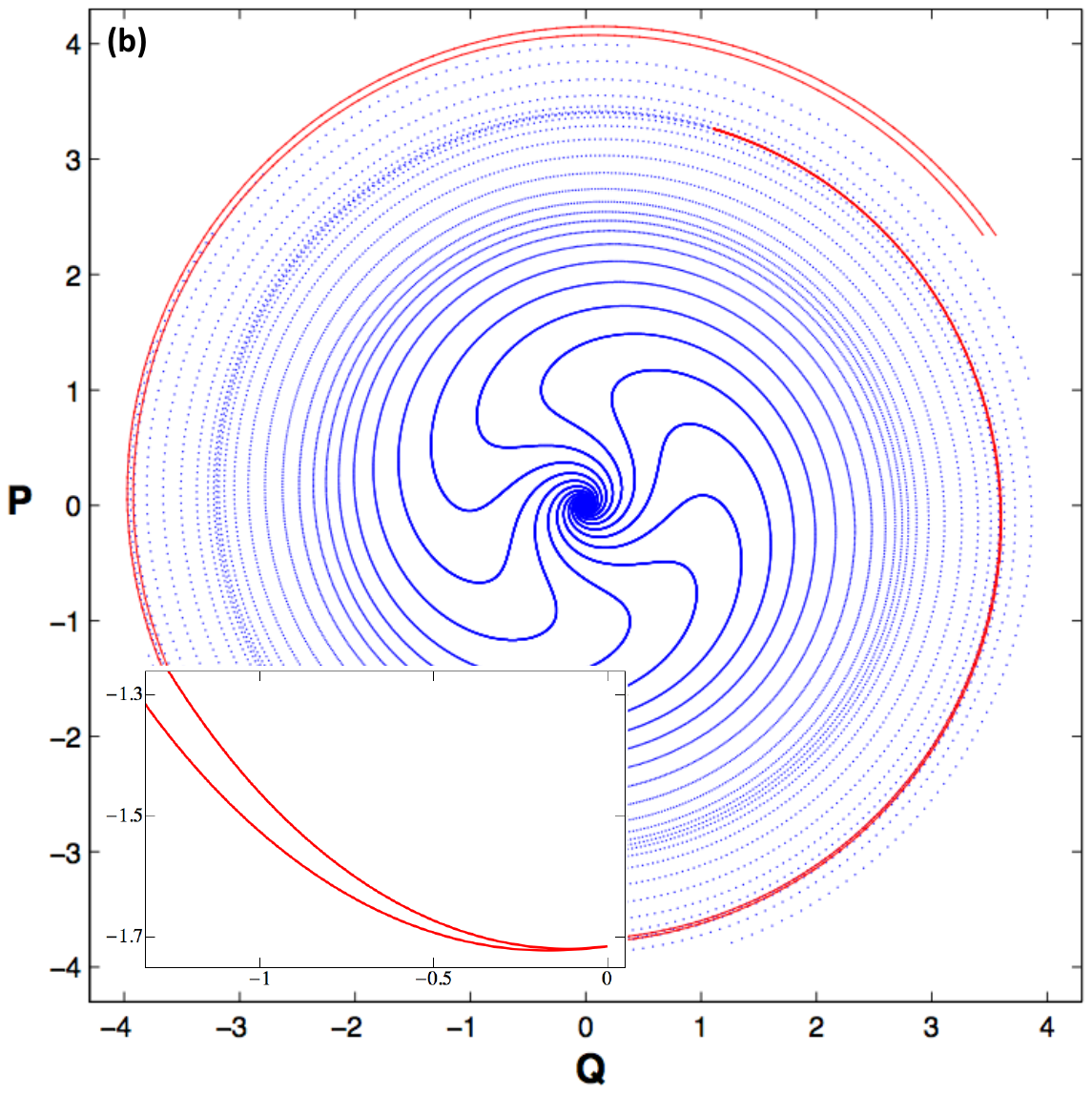}
\end{center}
\vspace{-1cm}
\caption{\label{fig:CombinedF} (a): The pattern of 10 escape trajectories
at $\mathcal{F} = 0$, $\eta =0.4$. The slow-down radius is $1$.  (b): The pattern of 10 escape trajectories at
$\mathcal{F} = 0.008$, $\eta =0.4$.  Increased density of trajectories is clearly seen.  The trajectories intersect along a caustic, shown in red, which originates at the cusp.  Due to the spiral nature of the trajectories, their intersections can not be easily seen by eye - the caustic has been mapped out with the help of the technique outlined in Secion \ref{sec:num_proc}.  (b)-inset:  As $\mathcal{F}$ is increased the caustic system evolves and the cusp becomes more open.  Here the small portion of the caustic around the cusp point is shown for $\mathcal{F} = 0.05$, $\eta =0.4$.  At this value of $\mathcal{F}$ the caustic system becomes very complicated.  The full details of it are not shown, and are not expected to be generic;  only the cusp part of the caustic is depicted.    }
\end{figure}

All calculations of trajectories are terminated
either when a stopping radius is reached or when a trajectory
reaches a caustic.  In section \ref{sec:Analytics_linear} we show that the
radius of a trajectory at $\mathcal{F} = 0$ grows exponentially in time, so to provide an accurate solution valid at larger radii,
the time steps must decrease dramatically.  The choice of the
stopping radius was dictated mostly by the position of
features, such as caustics - if they are present; otherwise, as is the the case in Fig.~\ref{fig:CombinedF}-(a), a convenient stoping radius was chosen. 
For a given value of the stopping radius, the size of time step was decreased until the results
did not change significantly.  Fourth-order Runge-Kutta numerical integrator was used.

\begin{figure}
\begin{center}
\includegraphics[width=10cm]{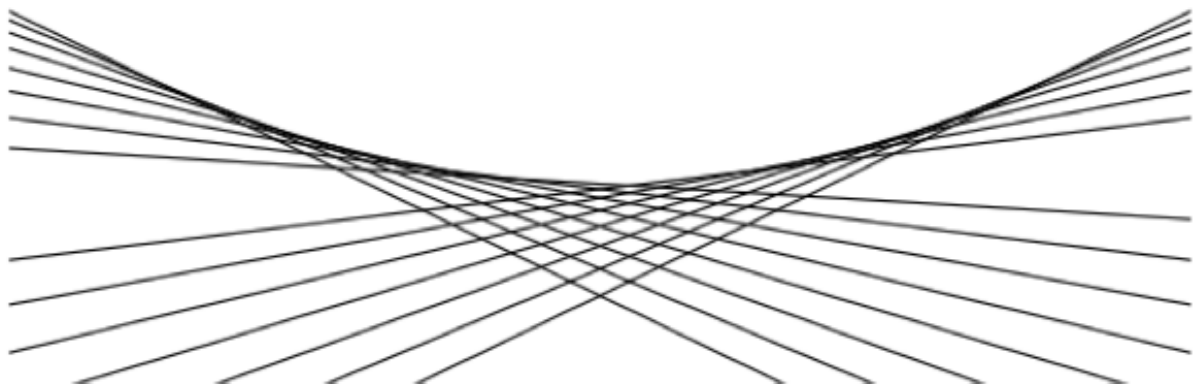}
\vspace{-5cm}
\caption{\label{fig:Cartoon} An envelope of intersecting neighboring straight lines is a simple example of caustic, as seen at the top of the figure.}
\end{center}
\end{figure}

A caustic is a the set of points where neighboring trajectories intersect - Fig.~\ref{fig:Cartoon}.
In our case we are concerned with the intersection of projections of trajectories of the auxiliary Hamiltonian dynamics.  Intersection of trajectories are of course an artifact of the projection - the evolution in the 4-dimensional auxiliary phase space with coordinates $(Q,P, p_Q, p_P)$ is smooth and unique.  As we mentioned, the trajectories in question evolve on the 2-dimensional Lagrangian manifold, which is a surface $\mathbf{p(r)}$, i.e., $(p_Q(Q,P), p_P(Q,P))$, which emanates from the fixed point.  In the vicinity of the fixed point, this manifold is a plane, but further away it may have folds.   As two neighboring trajectories pass over this fold, their projections will intersect; excellent cartoons of this can be found in \cite{Mark and Marko} and also in the beginning of \cite{Maier_Stein_2000}.  The fold is characterized by such location
where the tangent plane to the surface is perpendicular to the
$(Q,P)$ plane.  This happens when
\begin{equation}
\label{eq:Jacobean}
J = \frac{\partial(p_Q,p_P)}{\partial(Q,P)} = \infty,
\end{equation}
so caustics are mapped out from the knowledge of $\mathbf{p(r)}$.  

The function $\mathbf{p(r)}$ is not
known over all $\mathbf{r}$ - only $\mathbf{p}$ along a given trajectory can be calculated. However, as the
equations of motion for $p_Q$, $p_P$, $Q$ and $P$ are integrated
in time, it is also possible to simultaneously integrate the
equations of motion for second derivatives such as
$\frac{\partial^2 S}{\partial Q
\partial P} = \frac{\partial p_Q}{\partial P}$, etc. These the
time-dependent Riccati equations, 
\cite{MS SIAM}.
\begin{eqnarray}
\dot{S}_{ij} = -\frac{\partial^2 \mathcal{H}}{\partial p_k
\partial p_l}S_{ik}S_{jl} - \frac{\partial^2 \mathcal{H}}{\partial
r_j \partial p_k}S_{ik} - \frac{\partial^2 \mathcal{H}}{\partial
r_i \partial p_k}S_{jk} - \frac{\partial^2 \mathcal{H}}{\partial
r_i \partial r_j}. \nonumber \\
\end{eqnarray}
With the help of these and
the other four dynamical equations of motion we can track the evolution of $J$: all
together, there are 7 coupled ordinary differential equations. The initial conditions
are obtained from the knowledge of the surface $\mathbf{p(r)}$ in
the vicinity of the attractor, defined by unstable eigenvectors.
\subsubsection{Optimal fluctuational trajectories - numerical results}
\begin{figure}
\begin{center}
\includegraphics[width=10cm]{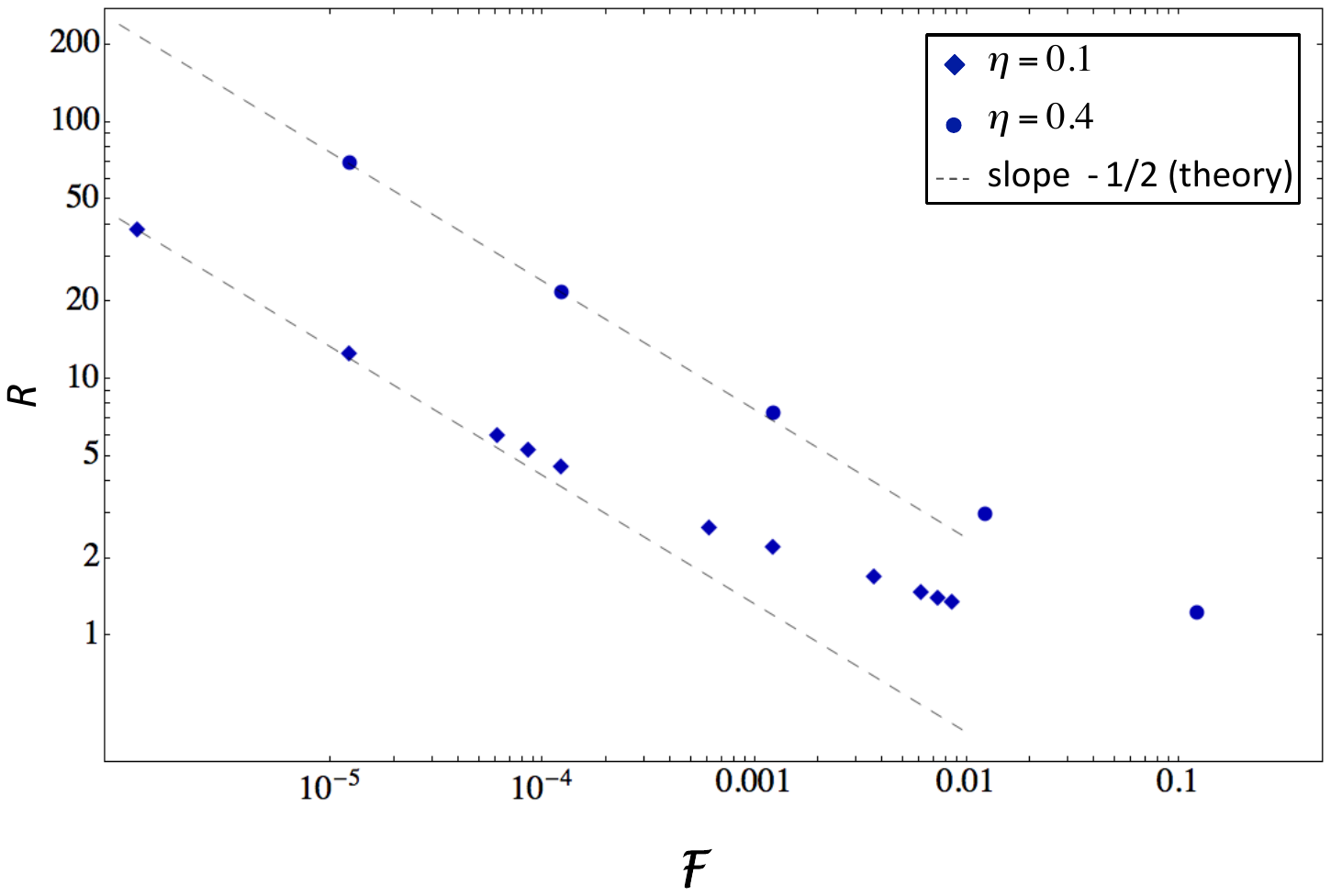}
\vspace{-2cm}
\caption{\label{fig:CuspScaling} Log-log plot representing the
radius of the cusp, measured from the origin versus the strength
of the driving $\mathcal{F}$ for $\eta = 0.4$ circles and $\eta =
0.1$ diamonds. The dashed lines represent power laws with exponent $-1/2$, a theoretical prediction outlined in the next section.}
\end{center}
\end{figure}

In this subsection we will study the pattern of fluctuational trajectories at ever
increasing driving strength $\mathcal{F}$ to elucidate the physics.  
When $\mathcal{F} = 0$, the pattern is rotationally symmetric, Fig.~\ref{fig:CombinedF}-(a), a fact that will be proven below.  As $\mathcal{F}$ becomes non-zero, the rotational symmetry of the pattern of trajectories becomes broken.  The density of trajectories also no longer obeys radial symmetry.  If an observation is made at a given radius, one would notice an ever increasing density of trajectories in certain regions until they actually ``jam'' into each other, and subsequently intersect.  This scenario is depicted in Fig.~\ref{fig:CombinedF}-(b), along with the numerically mapped location of the caustic, obtained by the method described Section \ref{sec:num_proc}.
%
The caustic displays a cusp structure, which is expected on the generic basis
(see \cite{Mark and Marko}, the Appendix 12 of \cite{Arnold} and
\cite{catastrophe_theory1, catastrophe_theory2}). As $\mathcal{F}$ is increased further, the cusp
continues to pull in closer to the origin and also becomes more open.  
We summarize the relationship between the radius of
the cusp and $\mathcal{F}$ in Fig.~\ref{fig:CuspScaling} for two values of $\eta$.   In this log-log plot we also show that each of the numerical sets of data is well-approximated by a straight line with slope $-1/2$, a power law that we predict analytically in section \ref{sec:Analytics_linear}.  As $\mathcal{F}$ continues to increase, the caustics evolve to become more complex.  We will not focus on the details of this in current work - our goal here is to focus on those features which are expected to be generic, such as the scaling properties of singularities as $\mathcal{F} \rightarrow 0$ and to answer the question whether there is a finite threshold in $\mathcal{F}$ at which the caustics appear.

\begin{figure}
\vspace{-0.3cm}
\begin{center}
\includegraphics[width=10cm]{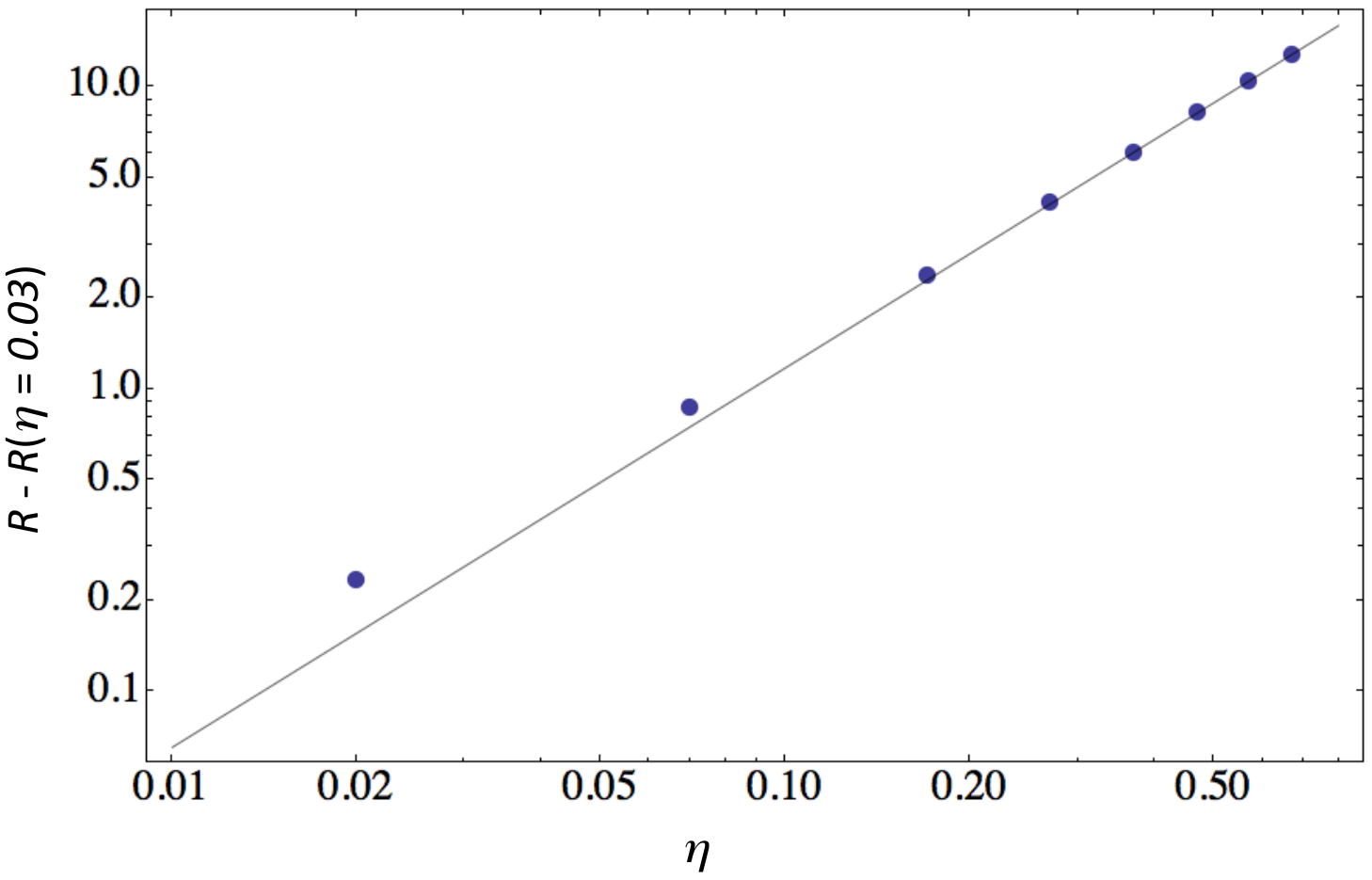}
\end{center}
\vspace{-2cm}
\caption{\label{fig:CuspScaling_versus_eta} Radius of the cusp
versus $\eta$ for a fixed $\mathcal{F} \approx 0.0012$.  The
asymptotic exponent $\tilde{p}$ in $R = 1+A\eta^{\tilde{p}}$ is
$\approx 1.26$.}
\end{figure}

Varying $\eta$ at fixed $\mathcal{F}$ also reveals an asymptotic scaling, 
Fig.~\ref{fig:CuspScaling_versus_eta}.
For the 
value of $\mathcal{F}$ shown, the relationship of $R$ versus $\eta$ becomes asymptotic
to a power law for larger $\eta$,
\begin{equation}
R = 1+A\eta^{\tilde{p}},
\end{equation}
where the exponent $\tilde{p} \approx 1.26$ for this value of $\mathcal{F}$. 
Moreover, since $\eta = \frac{\Gamma}{\delta \omega}$, where $\delta \omega$ is the deviation of the driving frequency
from the resonant frequency, 
we also learn that $R\left(\delta \omega\right)$ is
asymptotic to a power law near resonance.
\newpage \noindent
\subsection{Nonlinear Damping}
We can also consider the nonlinear interaction with the bath.  The Langevin description of this process is 
\begin{equation}
\label{eq:Langevin_nonlinear}
\ddot{q} + \gamma \dot{q} + \tilde{\gamma} q^2 \dot{q} + \omega_0^2 q + \alpha q^3 = F\cos{\left(\Omega t\right)} + f(t) + q \tilde{f}(t)
\end{equation}
with 
\begin{equation}
\langle \tilde{f}(t)\tilde{f}(t') \rangle = 2 \tilde{\gamma} k_B\tau  \delta(t-t').
\end{equation}
The corresponding slow-time amplitude equations can be put into the form \cite{note}
\small
\begin{eqnarray}
\label{eq:AmpEq_nonlinearQ}
\frac{d Q}{dT} &=& K_Q(Q,P)  +  f_Q(T) + Q\tilde{f}_Q(T) - P\tilde{f}_P(T) \\ 
\label{eq:AmpEq_nonlinearP} 
\frac{d P}{dT} &=& K_P(Q,P) + f_P(T) - P \tilde{f}_Q(T)  - Q\tilde{f}_P(T)
\end{eqnarray}
\normalsize
where 
\begin{eqnarray}
K_Q &=& \frac{\partial g}{\partial P} - \eta Q - \mu Q\left(Q^2 + P^2\right) \\
K_P &=& -\frac{\partial g}{\partial Q} - \eta P - \mu P\left( Q^2 + P^2 \right) \\
\label{eq:SlowNoise_CorrFn_nonlinear}
\langle \tilde{f}_i(T)\tilde{f}_j(T') \rangle &=& \tilde{D} \delta_{i,j} \delta(T-T'),
\end{eqnarray}
$\mu$ is defined in the Appendix, and 
$D/\tilde{D} = \eta/\mu$.
The action $S$ now contains new nonlinear damping and noise terms associated with nonlinear interaction of the oscillator with the bath
\small
\begin{multline}
\label{eq:S_nonlinear}
S  = \frac{1}{4} \int_{-\infty}^{\infty} \left(f_Q^2(t) + f_P^2 (t)\right)  \,dt + \frac{1}{4}\frac{\eta}{\mu}\int_{-\infty}^{\infty} \left(\tilde{f}^2_Q(t) + \tilde{f}^2_P (t)\right)  \,dt  \\
- \int_{-\infty}^{\infty} i \lambda_Q(t)\left(\dot{Q} - K_Q - f_Q(t) - Q \tilde{f}_Q(t) +  P \tilde{f}_P(t) \right) \,dt  \\
- \int_{-\infty}^{\infty} i \lambda_P(t)\left(\dot{P} - K_P - f_P(t) + Q\tilde{f}_P(t) + P\tilde{f}_Q(t) \right) \,dt   + O(D)
\end{multline}
\normalsize
The corresponding Wentzel-Freidlin Hamiltonian is
\begin{equation}
\mathcal{H} = \left(p_Q^2 + p_P^2\right)\left[1 + \frac{\mu}{\eta}\left(Q^2 + P^2\right)\right] + p_QK_Q + p_P K_P.
\end{equation}
\subsubsection{Optimal fluctuational trajectories - numerical results}
\begin{figure}
\begin{center}
\includegraphics[width=9cm]{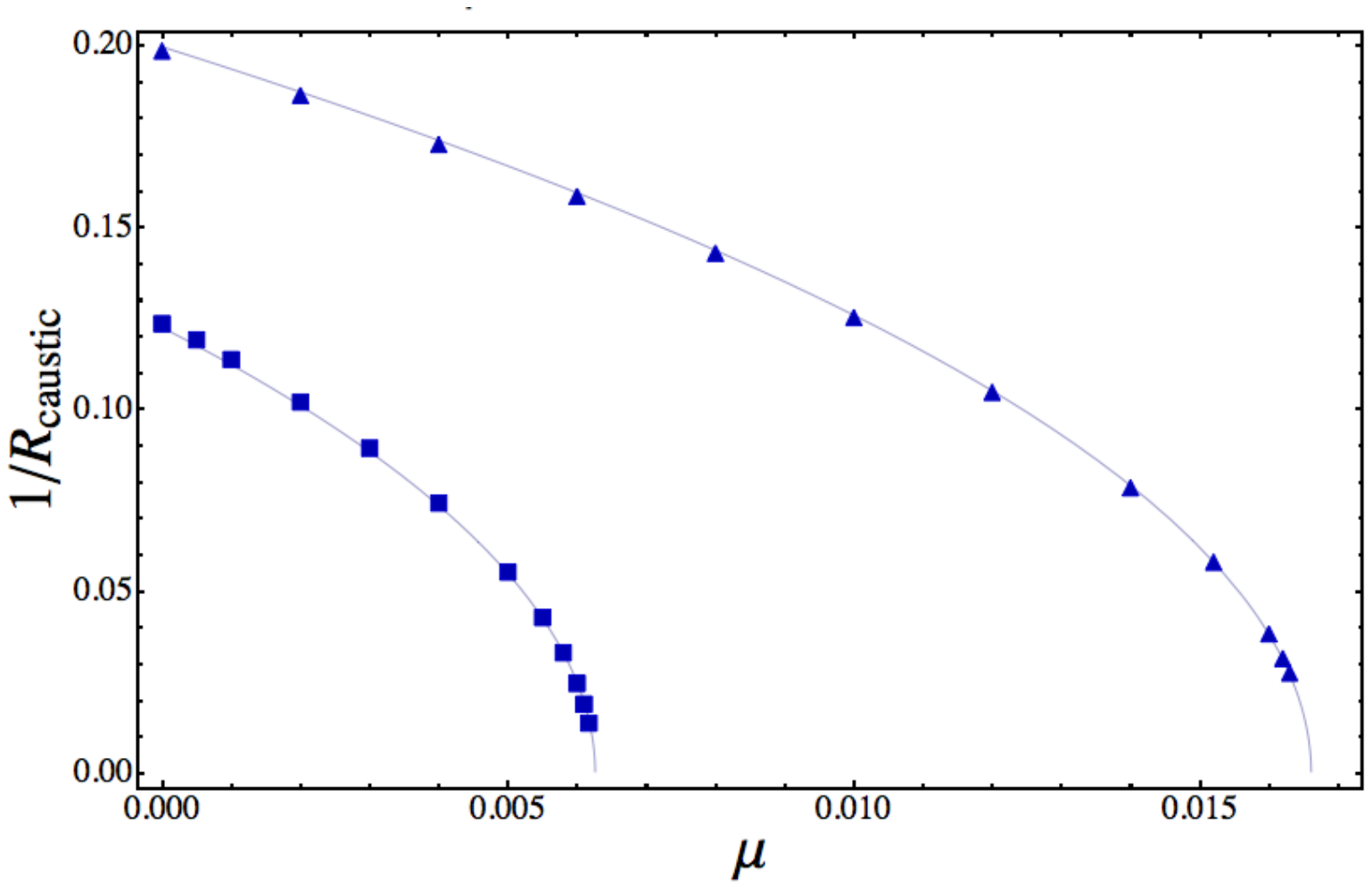}
\end{center}
\vspace{-1.8cm}
\caption{\label{fig:InvR_vs_mu} Inverse of the caustic radius $R$ versus the strength of the nonlinear damping $\mu$ for two different values of the driving force $\mathcal{F}$.  Plotting $1/R$ helps to see the existence of the limiting value of $\mu$ above which caustics disappear.  Solid curve - fit to the form $(\mu_{thr} - \mu)^{1/2}$.}
\end{figure}

\begin{figure}
\begin{center}
\includegraphics[width=9cm]{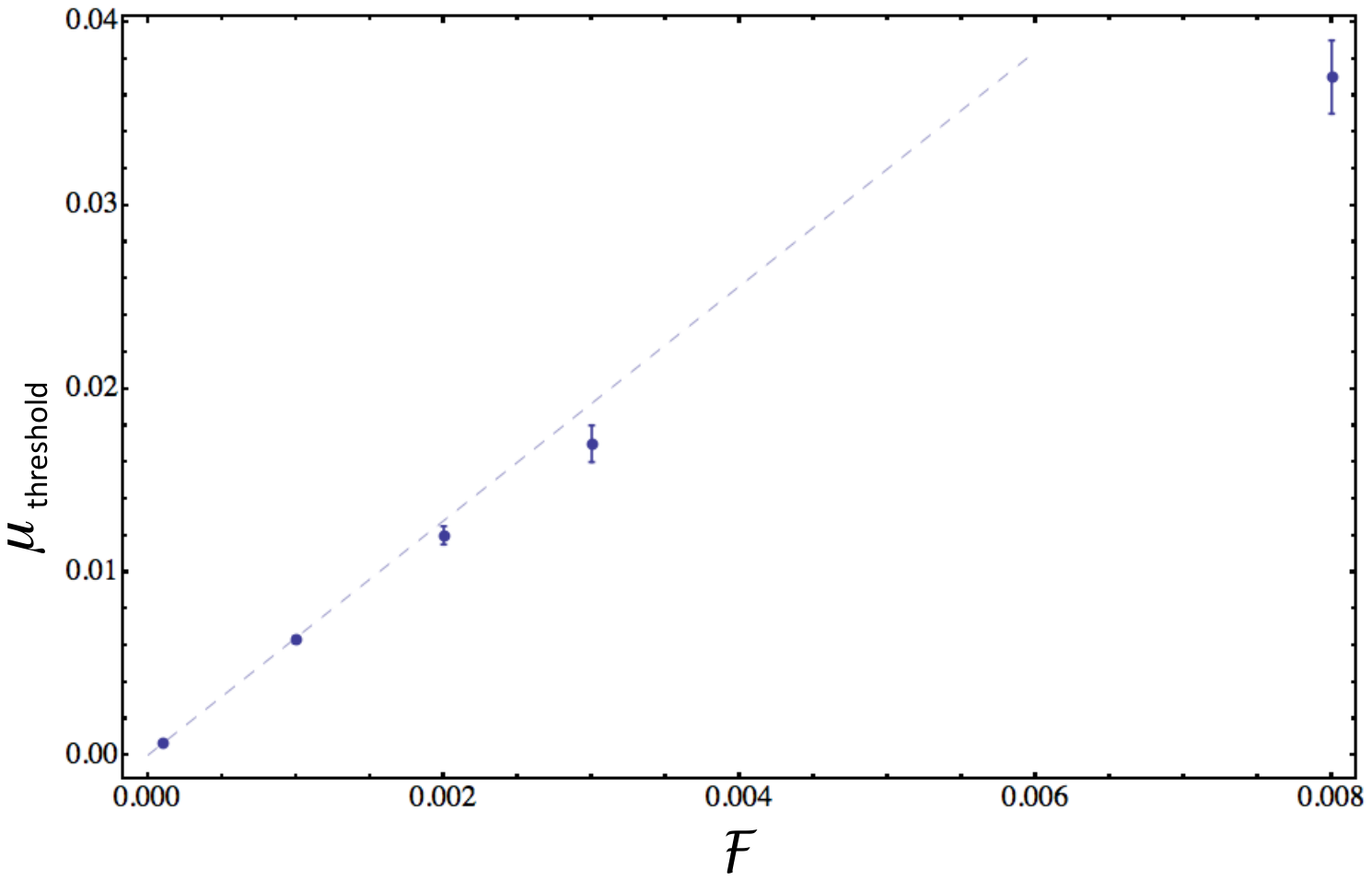}
\end{center}
\vspace{-1.7cm}
\caption{\label{fig:muthr_vs_F} Numerically-obtained $\mathcal{\mu}_{thr}$ verus $\mathcal{F}$ (dots).  At a given $\mathcal{F}$, the caustics appear for $\mu < \mu_{thr}$.  Alternatively, at a given $\mu$, the caustics appear for $\mathcal{F} > \mathcal{F}_{thr}$.  The error bars are obtained from the uncertainty of exact value of the threshold, as seen in Fig.~\ref{fig:InvR_vs_mu}.  The dotted line has a slope $6.4$.}
\end{figure}

We now present our numerical findings pertaining to the case of non-zero $\mu$.   The key observation is the extinguishing of caustics at large-enough values of $\mu$.  This is depicted in Fig.~\ref{fig:InvR_vs_mu}, where we plot the inverse of the radius of caustic versus $\mu$ for two different values of $\mathcal{F}$.  The $\mathcal{F}$-dependence at zero $\mu$ was depicted in Fig.~\ref{fig:CuspScaling}.  The data is very well approximated by a fit of the form 
\begin{equation}
\label{eq:R_vs_F_nlin_fit}
R_{caustic} \sim \left(\mu_{thr} - \mu\right)^{-1/2},
\end{equation}
strongly suggesting the existence of a critical value of nonlinear damping $\mu_{thr} \left( \mathcal{F} \right)$, above which the caustics are pushed away to infinity.  The existence of the such critical $\mu$ can also be inferred from the trend of the data of $R_{causitc}^{-1}$ versus $\mu$, as seen in Fig.~\ref{fig:InvR_vs_mu}.  
The quantity $\mathcal{\mu}_{thr}$ versus $\mathcal{F}$ is depicted in Fig.~\ref{fig:muthr_vs_F} for low values of $\mu$ and $\mathcal{F}$.  For a given value of $\mathcal{F}$ there is a critical value of $\mu$ below which caustics exist, $\mu_{thr}(\mathcal{F})$.  This $\mu_{thr}$ appears to scale linearly with $\mathcal{F}$ at low $\mathcal{F}$
\begin{equation}
\label{eq:muthr_vs_F_limit}
\lim_{\mathcal{F} \rightarrow 0} \mu_{thr}(\mathcal{F}) \sim \mathcal{F}.
\end{equation}
Equivalently, for a given value of $\mu$, a finite value of $\mathcal{F}$ is required for the formation of singularities, $\mathcal{F}_{thr}(\mu)$.   This, $\mathcal{F}_{thr}$ appears to scale linearly with $\mu$ at low $\mu$: 
\begin{equation}
\label{eq:Fthr_vs_mu_limit}
\lim_{\mu \rightarrow 0} \mathcal{F}_{thr}(\mu) \sim \mu.
\end{equation}
Thus,
\begin{equation}
R_{caustic} = \left(\mathcal{F} - \mathcal{F}_{thr}\right)^{-1/2}. 
\end{equation}
%
%
\section{Analytical Scaling Arguments}
\label{sec:Analytics}
The goal of the analytical treatment is to (i) explain various results of the previous section, but more importantly (ii) develop an intuitive understanding of the mechanism of the development of singularities.  This should serve our ultimate goal - to understand what properties of the system control the singularity threshold value of the tuning parameter that takes the system away from equilibrium.  The study of optimal paths will be based on the perturbation around the optimal path at zero $\mathcal{F}$, which we know from general arguments of statistical physics and can also calculate exactly.  The perturbation theory can not be used directly to identify the location of the caustic because the effect of the perturbation is always large on the caustic \cite{Schulman}; the weakness of the field is manifested in the fact that the caustic itself is pushed into far regions of the phase space.  However, quantities which obey power laws, such as the characteristic location of the caustic versus the strength of the driving field are nevertheless amenable to perturbative estimates.  In doing this, we extract the characteristic location of the caustic indirectly, by identifying those regions of the phase space where the zeroth-order term in the action becomes comparable to the first-order correction due to the driving.  Using this technique we were able to reproduce several scaling laws with respect to the driving field and with respect to the strength of the nonlinear damping.
\subsection{Linear Damping}
\label{sec:Analytics_linear}
In the limit of small noise the optimal path that minimizes $S$ dominates the functional integral, and a quadratic-order WKB approximation may be used (with some exceptions, such as near caustics \cite{Schulman}).  Applying the functional derivative to $S$,  we generate the following set of Euler-Lagrange equations for the optimal trajectory 
\begin{eqnarray}
\label{eq:linear_optimal-1}  f_Q &=& \dot{Q} - K_Q (Q(t),P(t))   \\
\label{eq:linear_optimal-2}  f_P &=& \dot{P} - K_P (Q(t),P(t)) \\
\label{eq:linear_optimal-3}  f_Q &=& -2i \lambda_Q \\
\label{eq:linear_optimal-4}  f_P &=& -2i \lambda_P \\
\label{eq:linear_optimal-5}  \dot{\lambda}_Q &=& -\lambda_Q \frac{\partial K_Q}{\partial Q} - \lambda_P \frac{\partial K_P}{\partial Q} \\
\label{eq:linear_optimal-6}  \dot{\lambda}_P &=& -\lambda_Q \frac{\partial K_Q}{\partial P} - \lambda_P \frac{\partial K_P}{\partial P}.
\end{eqnarray}
Using this it is easy to verify that at zero driving the optimal noise paths satisfy 
\begin{eqnarray}
\label{eq:linear_optimal-7} f^{(0)}_Q &=& 2\eta Q^{(0)} \\
\label{eq:linear_optimal-8} f^{(0)}_P &=& 2\eta P^{(0)}
\end{eqnarray}
from which it would also follow that
\begin{eqnarray}
\label{eq:linear_optimal-9}  \dot{Q} = \frac{\partial g}{\partial P} + \eta Q \\
\label{eq:linear_optimal-10}\dot{P} = -\frac{\partial g}{\partial Q} + \eta P.
\end{eqnarray}
These equations also follow from applying the transformation $T \rightarrow -T$, $Q \rightarrow Q$ and $P \rightarrow -P$ to the original mean-field equations (Eqs.~(\ref{eq:AmpEq_linearQ}) - (\ref{eq:g}) without noise), but this is just a time reversal relationship, as can be verified from the original Duffing equation (\ref{eq:Langevin_linear}).  The same would be accomplished by reversal the sign of $\eta$.   In other words, optimal fluctuations are time-reversed mean-field relaxations.  This is in accord with the general ``Principle of Maximal Work'' \cite{LL} for equilibrium systems.  

In view of Eqs.~(\ref{eq:S_linear}) and (\ref{eq:linear_optimal-1})-(\ref{eq:linear_optimal-2}), the full expression for the action reduces to
\begin{equation}
S =  \frac{1}{4} \int_{-\infty}^{0} \left(f_Q^2(t) + f_P^2 (t)\right) \,dt,
\end{equation}
although to evaluate this integral we need a full expression for $f_Q$ and $f_P$, which we do not know apriori.  We note however, that at zeroth-order we can use Eq.~(\ref{eq:linear_optimal-7}) and Eq.~(\ref{eq:linear_optimal-8})
to obtain
\begin{equation}
\label{eq:S^{(0)}-general-linear}
S^{(0)} = \eta^2 \int_{-\infty}^{0} \left[R^{(0)}(t)\right]^2 \,dt
\end{equation}
The solution to Eqs.~(\ref{eq:linear_optimal-9}) -(\ref{eq:linear_optimal-10}) for $P^{(0)}(t)$ and $Q^{(0)}(t)$ is most conveniently expressed in polar variables: 
\begin{eqnarray}
\label{eq: unperturbed_path_1}
R^{(0)}(t) &=& r e^{\eta t} \\
\label{eq: unperturbed_path_2}
\phi^{(0)}(t) &=& \theta + t - r^2 \frac{\left(e^{2\eta t} -1\right)}{2\eta}
\end{eqnarray}
Here $r$ and $\theta$ are initial conditions, to which a trajectory arrives at $t=0$ (hence the change in the upper limit of integration).  Combining this result for $R^{(0)}(t)$ with Eq.~(\ref{eq:S^{(0)}-general-linear}) we get
\begin{equation}
\label{eq:S_0_linear}
S^{(0)}(r,\theta) = \frac{\eta r^2}{2},
\end{equation}
where $(r,\theta)$ now play the role of an independent variable.  We mention in passing that one can also use the Wentzel-Freidlin action, Eq.~(\ref{eq:WF_action}) as the starting form for the analysis.  For example, the time-reversal symmetry of the fluctuational and relaxational paths can also be proven using this form as the starting point \cite{Thesis}. 

\subsubsection{Analysis of the first-order perturbation}
We seek $S^{(1)}$ in the series
\begin{equation}
S = S^{(0)} + \mathcal{F}S^{(1)} + ... .
\end{equation}
Since first order variation of $L^{(0)}$ around the unperturbed path is zero by definition, the only possible first order in $\mathcal{F}$ correction to the action is 
\begin{equation}
\label{eq:S1-general}
S^{(1)} = \int_{-\infty}^{0} L^{(1)}\left(Q^{(0)}(t), P^{(0)}(t), ...\right)\,dt.
\end{equation}
In view of this and of Eqs.~(\ref{eq:linearKQ})-(\ref{eq:linearKP}) and Eq.~(\ref{eq:S_linear}) we have
\begin{equation}
\label{eq:S1-details} 
S^{(1)} =  -\int_{-\infty}^{0} \eta P^{(0)}(t) \,dt
\end{equation}
where we also made use of Eq.~(\ref{eq:linear_optimal-4}) and Eq.~(\ref{eq:linear_optimal-8}).
We now analyze this contribution.  With the help of Eqs.~(\ref{eq: unperturbed_path_1})-(\ref{eq: unperturbed_path_2}) we have
\begin{equation}
S^{(1)} (r,\theta) = -\int_{-\infty}^{0} \eta r e^{\eta t} \sin{\left(\theta + t - \frac{r^2}{2\eta} \left(e^{2\eta t} -1\right)\right)}\,dt
\end{equation}
Notice that the dependence on the field point $(r, \theta)$ has been put into the parameters of the integrand, allowing the limits of integration to be simple.  We can change variables to $x = i\kappa e^{2\eta t}$ and re-express the form of $S^{(1)}$ in a more compact form:
\begin{equation}
S^{(1)} = -\frac{r}{2i} \frac{e^{i \theta}e^{i \kappa}}{2\left(i \kappa \right)^z} \int_{0}^{i\kappa} x^{z-1} e^{-x} dx + c.c.
\end{equation}
where $z = \frac{i}{2\eta} + \frac{1}{2}$ and $\kappa = \frac{r^2}{2\eta}$.
When $r \gg \sqrt{\eta}$ the upper limit can be replaced by $\infty$ (notice that $z$ is independent of $r$, so as $r$ is increased but $\eta$ is held fixed, only the upper limit changes, but the structure of the integrand does not, justifying this replacement).  The integral is taken in the x-plane along the imaginary axis, but the integrand is an analytic function  in the right half-plane, so we may rotate the contour to go along the real x-axis.  Then the integral coincides with the definition of the $\Gamma$-function in the variable $z$.  
After substituting $\kappa$ and a bit of algebra we have,
\small
\begin{equation}
\label{eq:S1lin} 
S^{(1)} 
= -\frac{\sqrt{2\eta}}{4}e^{i\left(\theta +\frac{r^2}{2\eta} - \frac{3\pi}{4} - \frac{1}{2\eta}\ln{\left(\frac{r^2}{2\eta}\right)} \right)}e^{\frac{\pi}{4\eta}} \Gamma\left(\frac{i}{2\eta} + \frac{1}{2} \right) + c.c.
\end{equation}
\normalsize
We note that limit $\eta \rightarrow \infty$ is actually equivalent to starting out with the integral that does not contain the $\ln$-term. 
The $S^{(1)}$ just obtained is a non-growing function, but with ever increasing oscillation frequency in the radial variable.   Next, we  take a Jacobean, Eq. (\ref{eq:Jacobean}).  Using $p_i = \frac{\partial S}{\partial q_i}$, the series expansion for $S$, and Eq.~(\ref{eq:S_0_linear}), the Jacobean has a compact form
\begin{eqnarray}
\label{eq:J}
J &=& \eta^2 + \eta \mathcal{F}\nabla^2 S^{(1)} + O(\mathcal{F}^2)  \nonumber  \\
&\equiv& J^{(0)} + \mathcal{F}J^{(1)} + ...
\end{eqnarray}
After some algebra we see that at large radii,
\begin{equation}
\label{eq:J-answer-linear}
\left|J^{(1)}\right| \sim \frac{r^2}{\eta^{1/2}} \times e^{\frac{\pi}{4\eta}} \left|\Gamma\left(\frac{i}{2\eta} + \frac{1}{2} \right)\right|
\end{equation}
Recall that exact $J$ must diverge on caustics.  In the language of perturbation theory, however, $J$ is unlikely to diverge (clearly the perturbative solution to the action $S^{(0)} + S^{(1)}$ is smooth and can not contain singularities; in fact, the Gaussian WKB theory is not meant to work on caustics \cite{Schulman}).  Therefore, the signature of caustics in the language of first-order perturbation theory is not the divergence of $J$, but the point at which such theory begins to break down.  In other words, the signature of caustics is the condition $\left|J^{(0)}\right| \sim \left|J^{(1)}\right|$.  Thus, Eq.~(\ref{eq:J-answer-linear}) suggests that 
\begin{equation}
\label{eq:r-caustic_lin}
R_{caustic} = \frac{\eta^{5/4}}{\mathcal{F}^{1/2}} \times e^{\frac{-\pi}{8\eta}} \left|\Gamma\left(\frac{i}{2\eta} + \frac{1}{2} \right)\right|^{-1/2}
\end{equation}
In view of Stirling's approximation the last factor approaches $(2\pi)^{-1/4}$ in the limit of $\eta \rightarrow 0$ and $\pi^{-1/4}$ in the limit $\eta \rightarrow \infty$.  Numerical computations show that this factor remains very close to the former constant value until $\eta \sim 1$, after which there is a crossover into the latter value.  Therefore the square root factor in 
 Eq.~(\ref{eq:r-caustic_lin}) can be disregarded, and we recover the numerical result for scaling with respect to the driving field and friction, 
 \begin{equation}
 R_{caustic} \sim  \frac{\eta^{5/4}}{\mathcal{F}^{1/2}}.
 \end{equation}
 \subsection{Nonlinear Damping}
The analysis of \ref{sec:Linear_Damping_Analysis} can be repeated for the case of nonlinear damping.  As in the linear case, we can take variational derivatives (there are two more now, with respect to the new noises), and find optimal $f(t)$s, $\tilde{f}(t)$s, $\lambda(t)$s and $\left(Q(t), P(t) \right)$.  First, we find that at $\mathcal{F} = 0$ 
\begin{eqnarray}
\label{eq:nonlinear_optimal-3} \tilde{f}^{(0)}_Q &=& 2\mu \left(Q^2 - P^2 \right) \\
\label{eq:nonlinear_optimal-4} \tilde{f}^{(0)}_P &=& -4 \mu QP.
\end{eqnarray}
in addition to Eqs.~(\ref{eq:linear_optimal-7})-(\ref{eq:linear_optimal-8}), which once again imply that the optimal paths at zero driving are time-reversed relaxations.  Along the optimal path the full action is given by 
\begin{equation}
S = \frac{1}{4} \int_{-\infty}^{0} \left(f_Q^2(t) + f_P^2 (t)\right)  \,dt + \frac{1}{4}\frac{\eta}{\mu}\int_{-\infty}^{0} \left(\tilde{f}^2_Q(t) + \tilde{f}^2_P (t)\right).
\end{equation}
We can also find zeroth-order trajectories:
\begin{eqnarray}
\label{eq: unperturbed_path_1_nonlinear} R^{(0)}\left(t\right) &=&  \frac{re^{\eta t}}{\sqrt{1 + \frac{\mu r^2}{\eta} \left( 1- e^{2\eta t}\right)}}  \\
\label{eq: unperturbed_path_2_nonlinear} \phi^{(0)}\left(t\right) &=& \theta + t + \frac{1}{2\mu} \ln{\left(1 + \frac{\mu r^2}{\eta} \left(1 - e^{2\eta t}\right)\right)}
\end{eqnarray}
Combining all these results we get
\begin{equation}
\label{eq:S_0_nonlinear}
S^{(0)}(r,\theta) = \frac{\eta r^2}{2},
\end{equation}
same as in the linear damping case.  
\subsubsection{Analysis of the first-order perturbation}
We move on to calculate $S^{(1)}$ for the case with nonlinear damping.  The statement in Eq.~(\ref{eq:S1-general}) remains true.  Then, in view of Eqs.~(\ref{eq:AmpEq_nonlinearQ})-(\ref{eq:AmpEq_nonlinearP}) and Eq.~(\ref{eq:S_nonlinear})  we have, as in the linear case
\begin{eqnarray*}
S^{(1)} =  -\int_{-\infty}^{0} \eta P^{(0)}(t) \,dt,
\end{eqnarray*}
where we also made use of Eq.~(\ref{eq:linear_optimal-4}) and Eq.~(\ref{eq:linear_optimal-8}).   With the help of Eqs.~(\ref{eq: unperturbed_path_1_nonlinear})-(\ref{eq: unperturbed_path_2_nonlinear}) and 
changing variables to $\xi = e^{\eta t}$ gives
\begin{equation}
\label{eq:S1-nonlinear}
S^{(1)} 
= -\frac{re^{i \theta}}{2i}\int_0^1 \xi^{i/\eta}\left(1+\frac{\mu r^2}{\eta} \left(1-\xi^2\right)\right)^{\frac{i}{2\mu} - \frac{1}{2}} \,d\xi + c.c. 
\end{equation}
This is the general form.  
We simplify notation by introducing $\alpha = \frac{\mu r^2}{\eta}$ and $\Delta = 1+ \alpha$.  There are now two important regimes to consider: regime where the nonlinear damping dominates the linear damping, $\alpha \gg 1$, so $\frac{\alpha}{\Delta}\approx 1 - \frac{1}{\alpha} \rightarrow 1$ and the regime of perturbative nonlinear damping, where it is weak compared to the linear damping, $\alpha \ll 1$, so $\frac{\alpha}{\Delta} \approx \alpha \ll 1$.

In the first regime, the second term in the brackets in Eq.~(\ref{eq:S1-nonlinear}) dominates. 
This allows all the $r$-dependence to be taken out of the integral, giving 
\begin{equation}
S^{(1)} \sim r^{-\frac{i}{\mu}}
\end{equation}
We apply Eq.~(\ref{eq:J}) to take the Jacobean, and find that 
\begin{equation}
J^{(1)} \sim r^{-2}
\end{equation}
This explains the existence of a threshold in $\mathcal{F}$.  A sufficiently large $\mathcal{F}$ is necessary at any large $r$ to bring $\mathcal{F}J^{(1)}$ to be comparable to $J^{(0)}$ ($ = \eta^2$).  In contrast, the case of linear damping taught us that $J^{(1)}$ is an always increasing function ($\sim r^2$), so for any finite value of $\mathcal{F}$ there exists a radius where $\mathcal{F}J^{(1)}$ dominates $J^{(0)}$.  

We now consider the second mentioned regime. 
For this task, we rewrite Eq.~(\ref{eq:S1-nonlinear}) as follows
\begin{equation}
S^{(1)} = -\frac{re^{i\theta}}{2i} \Delta^{\frac{i}{2\mu} - \frac{1}{2}} \int_0^1 \xi^{i/\eta}\left(1-\frac{\alpha}{\Delta} \xi^2 \right)^{\frac{i}{2\mu} - \frac{1}{2}} \,d\xi + c.c.
\end{equation}
Analogously to the case with linear damping, now define a new variable $y = \kappa \xi^2$, where $\kappa = \frac{r^2}{2\eta}$.  Also notice that in the limit $\alpha \rightarrow 0$, we have $\alpha/\Delta \rightarrow \alpha$.  Then,
\begin{equation}
\label{eq:S1-nonlinear-small_mu}
S^{(1)} = -\frac{r e^{i \theta}}{4i} \Delta^{\frac{i}{2\mu} - \frac{1}{2}} \kappa^{-\frac{i}{2\eta} - \frac{1}{2}}\int_0^{\kappa} y^{\frac{i}{2\eta} - \frac{1}{2}} \left(1- 2\mu y   \right)^{\frac{i}{2\mu} - \frac{1}{2}} \,dy + c.c.
\end{equation}
%
%
As in the case of linear damping we assume:  $r \gg \sqrt{\eta}$, i.e. $\kappa \gg 1$.  The integral in Eq.~(\ref{eq:S1-nonlinear-small_mu}) is then a function of only $\mu$ and $\eta$, so it will not affect our quest for scaling with $r$.  Thus,
\begin{equation}
S^{(1)} = -e^{i \theta} e^{\frac{i}{2\mu} \ln{\left(1+\frac{\mu r^2}{\eta}\right)}} \left(1 - \frac{\mu r^2}{2\eta} \right) e^{-\frac{i}{2\eta}\ln{\left(\frac{r^2}{2\eta}\right)}}  f(\mu, \eta) + c.c. 
\end{equation}
where
\begin{equation}
\lim_{\mu \rightarrow 0} f(\mu, \eta) = \frac{\sqrt{2\eta}}{4} e^{\frac{\pi}{4\eta}} e^{-\frac{3i\pi}{4}}\Gamma\left(\frac{i}{2\eta} + \frac{1}{2} \right) + c.c.
\end{equation}
This $S^{(1)}$ agrees with Eq.~(\ref{eq:S1lin}) as $\mu \rightarrow 0$ (the identity $lim_{\epsilon \rightarrow 0} \left(1-\epsilon x\right)^{1/\epsilon} = e^{-x}$ may be helpful).  Expanding the logarithm in powers of $\mu$ and doing more algebra we find that a $\mu$-contribution to the Jacobean with the largest power in $r$ is of the form
\begin{equation}
-\mathcal{F} \mu\frac{r^4}{\eta^2} \times e^{i(...)}f_0(\eta).
\end{equation}
This is the leading-order addition to the linear damping Jacobean when $r$ is large (i.e. $r \gg \sqrt{\eta}$), while $\frac{\mu r^2}{\eta}$ is small.  In other words, as we go to large $r$ and decrease $\mu$ to be in the regime of perturbative effect of the nonlinear damping, the correction to the Jacobean will take the following asymptotic form [c.f. Eq.~(\ref{eq:J-answer-linear})]
\begin{equation}
\left|J^{(1)}\right| \sim   \left|\left(\frac{r^2}{\eta^{1/2}} - \mu\frac{r^4}{\eta^{3/2}}\right) \times\Gamma\left(-\frac{i}{2\eta} + \frac{1}{2} \right)\right|e^{\frac{\pi}{4\eta}}. 
\end{equation}
Comparing $\left|J^{(0)}\right|$ with $\left|J^{(1)}\right|$, and suppressing the $\eta$-dependence, we have
\begin{equation}
R_{caustic} \sim \frac{1}{\mathcal{F}^{1/2}} + \frac{\mu}{2\mathcal{F}^{3/2}}.
\end{equation}
Nonlinear damping pushes the caustic to a larger radius.  This describes the small-$\mu$ part of Fig.~\ref{fig:InvR_vs_mu}.  In fact, combining Eqs.~(\ref{eq:R_vs_F_nlin_fit}) - (\ref{eq:muthr_vs_F_limit}) and expanding it in series in $\mu$ (again suppressing the $\eta$-dependence) we reproduce this result.  However, the perturbation theory does not give prediction for an arbitrary value of $\mu$, such as empirically-obtained Eq.~(\ref{eq:R_vs_F_nlin_fit}).  

\section{Summary and Discussion}
\label{sec:Discussion}
We considered a simple system interacting with a heat bath.  The amplitude of the periodic driving $\mathcal{F}$ is the control parameter that measures the deviation of the system from thermal equilibrium.  We may summarise our findings as follows.  At $\mathcal{F} = 0$ the pattern of optimal fluctuational paths is smooth, and is related to the pattern of relaxational paths via the change of the signs of linear and nonlinear damping.   A unique fluctuational path leads to a given point $(Q,P)$ in the phase space in the stationary regime. 
In the case of linear damping, the optimal paths at $\mathcal{F} = 0$ have exponentially-growing radii, and hence reach infinity in infinite time.  With finite nonlinear damping $\mu$, the exponential growth is replaced by a faster growth beyond a certain radius, such that the optimal path reaches infinity in finite time.  This qualitative difference determines whether the driving threshold that is necessary for the appearance of singularities will be finite or infinitesimal.

In the case of $\mu = 0$, we observe the appearance of singularities for an infinitesimal value of $\mathcal{F}$.  
The scenario by which this happens is such that the singularities appear at infinity and move in to smaller radii as $\mathcal{F}$ is increased: $R_{caustic} \sim \mathcal{F}^{-1/2}$.   Interestingly, the singularities happen not where the rotation switches from clockwise to conter-clockwise motion, see Fig.~\ref{fig:CombinedF}; after all, the region where the angular part of the dynamics slows down is expected to be most sensitive to perturbations.  We learned that the mechanism for the formation of singularities is instead related to skewness, or uneven distortion of the family of optimal paths.  The smaller the $\mathcal{F}$, the smaller is this distortion, and the larger is the distance required for the distortion to cause neighboring paths to intersect.  To verify that the slowing-down region is non-essential for the appearance of caustics, we considered a system where $g$ is given by
\begin{equation}
g = \frac{1}{4}\left(Q^2 + P^2 + 1\right)^2 - Q\mathcal{F}
\end{equation}
(compare to Eq.~(\ref{eq:g})).  The optimal paths of this system 
rotate only in one direction and do not display a turn-around effect as do the paths we have studied so far.  However, we found the caustics in this system as well.

In the case when the nonlinear damping $\mu$ is non-zero, we found that
\begin{eqnarray}
R_{caustic} &\sim& 
\left(\mathcal{F} - \mathcal{F}_{thr}\right)^{-1/2} \\
\mathcal{F}_{thr} &\sim& \mu
\end{eqnarray}
Thus, for a fixed value of $\mu$, varying $\mathcal{F}$ has the effect of bringing the caustics in from infinity.  
We can understand the origin of the linear relationship $\mathcal{F}_{thr} \sim \mu$ from the following argument.  There are two effects.  On the one hand, a non-zero driving $\mathcal{F}$ introduces skewness of the pattern of optimal paths.   The characteristic radius at which the skewness is sufficient to create a singularity scales like $\mathcal{F}^{-1/2}$  at zero $\mu$.   On the other hand, a finite value of $\mu$ causes the optimal paths to accelerate; their dynamics switches from an exponentially growing to an even faster one, causing them to reach infinity in finite time.  
This change in the dynamics of optimal paths happens at a characteristic radius $\sim \left(\eta/\mu\right)^{1/2}$.  When the motion is faster than exponential at the radius where the skewness would otherwise cause the singularity, this fast growth prevents it.   Equating these two characteristic radii gives $\mathcal{F}_{thr} \sim \mu$.  As $\mathcal{F}$ is increased, the region where the skewness would cause the singularity moves to a smaller radius - into the exponentially-growing part of the radial dynamics, and caustics appear.  This is the picture of the singularity formation in a monostable system - skewness, rather than slowing-down of paths, is the key property responsible for the birth of singularities - as long as fluctuational paths move slowly-enough in the relevant part of the phase space.  Due to the fact that our system is minimal - it contains damping and nonlinearity and has only three control parameters, we believe this picture may be generic.

There is a unique correspondence between singularities in the pattern of optimal paths and the singularities in the probability distribution \cite{Mark and Marko}, as mentioned in the Introduction.  Indeed, it has been pointed out almost twenty years ago by Graham and T\'{e}l \cite{Graham DB} that if there exists a nonequilibrium potential, it will be generically non-smooth.  The skewness that is the key to our picture is a consequence of the loss of integrability -  at finite $\mathcal{F}$ the energy of the auxiliary dynamics is the only integral of motion, while at $\mathcal{F} = 0$ the angular momentum is also conserved.  One may ask if any loss of integrability will generically lead to singularities?   This work demonstrates that a loss of integrability is not always sufficient for formation of singularities - even when the auxiliary dynamics is non-integrable, the nonequilibrium potential is everywhere smooth if the driving strength is below a certain threshold value. This value depends on the speed of the mean-field dynamics at equilibrium.   


The results of this work have been obtained with M. I. Dykman, and will be submitted to a peer-reviewed journal jointly.  The author thanks Michael Cross and Michael Khasin for useful discussions; Carl Goodrich, Tony Lee and Eyal Kenig for very helpful proof-reading suggestions, and the NSF grant number DMR-0314069 for support in the early stages of this work.

\appendix
\section{Dimensionless variables}
Variables $(Q,P)$ are related to $(q,\dot{q})$ via a canonical transformation:
\small
\begin{equation}
q \equiv \frac{\omega_0}{\alpha^{1/2}} \sqrt{\frac{8}{3}} \sqrt{\frac{|\Omega - \omega_0|}{\omega_0}} \left[P\sin{(\Omega t)} + Q\cos{(\Omega t)} \right] 
\end{equation}
\begin{equation}
\frac{dq}{dt} \equiv \frac{\omega_0 \Omega}{\alpha^{1/2}} \sqrt{\frac{8}{3}} \sqrt{\frac{|\Omega - \omega_0|}{\omega_0}} \left[P\cos{(\Omega t)} - Q\sin{(\Omega t)} \right]
\end{equation}
\normalsize
If a complex convention is used (as in \cite{Mark Review}):  $q \equiv ue^{i\Omega t} + c.c.$ and $\dot{q} \equiv i\Omega u e^{i\Omega t} + c.c.$, the resulting amplitude equations are related to the present ones by the relationship $P \rightarrow -P$.  The rest of the dimensionless variables and parameters are related to the physical ones as follows:
\begin{eqnarray}
\eta &=& \frac{\gamma}{2|\Omega - \omega_0|}  \\
\mathcal{F} &=& \frac{\sqrt{3}}{2^{5/2}}\frac{1}{|\Omega - \omega_0|^{3/2}}\frac{\sqrt{\alpha}}{\omega_0^{3/2}} F  \\
D &=&  \frac{3\alpha}{8 \omega_0^3 \left(\Omega - \omega_0\right)^2} \gamma k_B\tau \\
T &=& |\Omega - \omega_0|t = \gamma t \\
\mu &=& \frac{\tilde{\gamma}\omega_0}{3\alpha}
\end{eqnarray}


\begin{thebibliography}{50}
\bibitem{note} This 
can also be obtained from the microscopic Hamiltonian with a nonlinear interaction between the bath degrees of freedom and the oscillator, following \cite{Mark Review}.
\end{thebibliography}
\end{document}